\documentclass[conference]{IEEEtran}
\IEEEoverridecommandlockouts
\usepackage{cite}
\usepackage{amsmath,amssymb,amsfonts}
\usepackage{algorithmic}
\usepackage{graphicx}
\usepackage{textcomp}
\usepackage{xcolor}
\usepackage[numbers]{natbib}

\usepackage{subfigure}

\def\BibTeX{{\rm B\kern-.05em{\sc i\kern-.025em b}\kern-.08em
    T\kern-.1667em\lower.7ex\hbox{E}\kern-.125emX}}
\begin{document}

\title{Adaptive Data Path Selection for Durable Transaction in GPU Persistent Memory
}


\author{\IEEEauthorblockN{1\textsuperscript{st} Xinjian Long, 
}
\IEEEauthorblockA{\textit{State Key Laboratory of Networking and Switching Technologies} \\
\textit{Beijing University of Posts and Telecommunications}\\
Beijing, China \\
barbiel\_origin@bupt.edu.cn
}

}

\maketitle

\begin{abstract}
The new non-volatile memory technology relies on data recoverability to achieve the promise of byte-addressable persistence in computer applications. The durable transaction (e.g. logging) is one of the major persistency programming models to provide recoverable data structures. To achieve performant failure-atomic transactional updates to PM, multi-data-path architectures that separate the data paths for persists are recently explored for CPUs. Considering the importance of GPU as a key computing platform for many application domains, we investigate the multi-data-path architecture for durable transactions to PM in GPU. Our solution, AGPM, exploits an adaptative data-path-selection strategy for the log updates to PM. AGPM reduces the GPU kernels' execution time by at least 24.37\% (at most 66.44\%) compared to the state-of-the-art designs.
\end{abstract}


\section{Introduction}


Persistent memory (PM) or non-volatile memory (NVM) technologies have received significant attention from both academia \cite{ref1,ref2,ref3,ref4,ref5,ref6,ref7} and industry \cite{ref8,ref9,ref10,ref11}. PM is expected to provide high integration density comparable to disk storage at latencies comparable to DRAM \cite{ref12} while supporting byte-grained addressability and durability \cite{ref13}. The recent commercialization of PM is Intel's Optane NVM \cite{ref14}. Years of PM research on the CPU have been stacked \cite{ref15,ref16,ref17,ref18,ref19,ref20,ref6}, but the corresponding exploration on the GPU is surprisingly limited, let alone the hardware implementation of PM onboard the modern GPU. However, features of PM (increased memory capacity, high access speed, data recoverability, etc.) are proven beneficial for GPU workloads \cite{ref21}. Thus, it is necessary for GPU to support PM, and there is a lot of potential for GPU-specific PM design that yields high performance.


Data recoverability is one of the key reasons why GPU applications are benefited from the PM. For instance, long-running GPU applications (deep neural network training, proof-of-work algorithms in blockchain applications, etc.) can obtain performance improvement by relying on recoverable data structures residing in the main memory instead of conducting costly system checkpointing \cite{ref22, ref23}. 
Persistency model \cite{ref16,ref7,ref24,ref25,ref6} is necessary for the implementation of the recoverable data structure. This is because the commit order of the PM updates could be different from the order they reach the PM due to the existence of the volatile write-back caches. Besides, persistency models can provide transaction-like semantics \cite{ref6,ref8,ref26}, which enables a part of the program, delineated by durable transaction, to be recoverable with all the data persisted therein all or nothing. 
Durable transactions could apply separate data paths to move data to PM for durability and consistency as needed. On the one hand, a $store$ followed by a $clwb$ instruction can be used to persist a specific line from the volatile cache hierarchy to PM (temporal path). On the other hand, applications can also use non-temporal instructions to bypass the cache and write directly to the PM (non-temporal path). Upon analysis of different GPU workloads, we observe that the inappropriate selection of the data path further deteriorates the performance problem. Most of the current persistency models are designed to statically use a single-data-path architecture, which risks the applications using PM experiencing unnecessary performance loss.


Recently, Shahri et al. \cite{ref27} proposed an extension to the x86 memory persistency model to exploit the differences in speeds of requests sent along the temporal and the non-temporal paths to reduce the reliance on using $sfence$ instruction. Jeong et al. \cite{ref28} proposed a hardware/software co-design scheme that leverages the separate FIFO data paths to enforce persisting orders. 
These designs decided to only consider the types of PM requests ($load$, $store$, etc.) as the key factor to selecting the data path. For example, when using undo logging, the log updates must persist before the corresponding data updates. In this case, logs are intuitively updated using the non-temporal path instead of the temporal one \cite{ref13,ref29}. This is because the non-temporal path is commonly faster than the temporal one by skipping the multiple levels of caches between the processor and the persistent domain. In addition, logs are expected to never be read during failure-free execution, and the usage of the temporal data path's write-no-allocate policy can avoid polluting the volatile caches while reducing the undo logging's overhead. In contrast with this intuition, we observe that the performance of using different data paths varies across GPU applications due to their different memory access behaviors. In some cases, we observe that log updates using the slow temporal path can be more beneficial than using the fast non-temporal path. 
On the other hand, the temporal path is not always more beneficial for data updates with temporal and spatial localities. Specifically, the non-temporal path is more recommended compared to the temporal one when the data locality is too strong and the reuse distances are too long to help the cached data be reused before eviction. 
Overall, our observation manifests that it is not amenable to handling GPU PM requests using a static path selection strategy, and there is a lot of potential for the multi-path architecture to achieve better performance for GPU workloads' persist ordering.


Building on our observations, we propose an adaptive approach (AGPM) to select the data path for the durable transactions in GPU. This approach adds AGPM buffers in the GPU memory hierarchy to record the data locality between the PM log updates and the subsequent data updates, as well as the locality among the log updates, within the GPU L1D caches and the shared L2 cache. Periodically, statistics in the AGPM buffers are fed to an added path selector in each SM and are translated to 8 reasons/patterns. Such translation is conducted upon the temporal and spatial locality of the PM logging data in the L1D and the L2 caches, the GPU kernel's usage of the shared memory, and the number of log updates. Reasons/patterns are exploited to help derive the data path selection decision for undo logging. Overall, this paper makes the following contributions:

\begin{enumerate}

\item We analyze the L1D and L2 data localities for GPU kernels' transactional interaction with PM (undo logging) and their resulting performance in different types of GPU workloads.
\item We propose an adaptive approach for PM requests' data path selection in GPU referred to as AGPM, such that the PM log updates' data paths are selected according to the kernel accesses' data locality and characteristics.
\item Experimental results show that AGPM achieves significant performance improvement over the static data path selection methods. Meanwhile, AGPM outperforms the state-of-the-art (SOTA) multi-data-path architectures to PM. Compared to the SOTA designs, AGPM reduces the GPU kernels' execution time by at least 24.37\% (at most 66.44\%).
\end{enumerate}


The remainder of this paper is organized as follows. Section \ref{sec:backg} presents the background. Section \ref{sec:moti} discusses the motivation of this study. Section \ref{sec:sol} describes the design of the proposed AGPM. Section \ref{sec:eval} compares the results of our proposed AGPM with the other methods. Section \ref{sec:related} discusses the related works, before providing concluding remarks in Section \ref{sec:conclu}.

\section{Background}
\label{sec:backg}
In this section, we review the current persistency models on CPUs and GPUs. We also introduce the GPU architecture and the programming model following the NVIDIA/CUDA terminology. It is worth noting that the techniques mentioned in this section, as well as our design described in the following sections, are adaptable to other GPU architecture besides NVIDIA.

\subsection{Memory persistency models}
Byte-addressable persistent memory provides a promising future for high-performance in-memory computing with recoverable data structures (RDS). However, due to the volatile write-back caches, the order of $load$ and $store$ requests arriving at the persistent memory can be different from their commit order. This breaks the rule of RDS. For example, assuming that $p$ is a persistent structure. '$p$ $\rightarrow$ data' and '$p$ $\rightarrow$ state' reside in different cache lines and the update to '$p$ $\rightarrow$ data' precedes the update to '$p$ $\rightarrow$ state' in the program. However, the '$p$ $\rightarrow$ state' in memory may be updated before '$p$ $\rightarrow$ data' due to the caches. In this case, if a fault (e.g., a power failure) happens, the persistent memory state becomes incorrect after power is restored. To deal with this issue and support correct implementations of RDS, memory persistency models \cite{ref16,ref7,ref24,ref25,ref6,ref30} are proposed to formally specify the order of writes to PM. These models can be broadly characterized as the strict and the relaxed models, which are distinct in the levels of concurrency. In the strict persistency model, the order of the volatile memory operations is identical to the order of the PM operations. This model is easy to implement, but it limits the program's concurrency which leads to the most serious performance degradation. The relaxed persistency model is more performant by breaking the tie between the volatile memory operations and the PM operations. A higher level of PM writes concurrency is supported at the cost of program annotation and hardware complexities.

The memory persistency models mentioned above specify the durability order of stores but ordering alone does not guarantee data recoverability. For example, assume that with either a strict or relaxed persistency model, '$p$ $\rightarrow$ data' and '$p$ $\rightarrow$ state' in PM are updated in the program order. But it is still possible that a fault happens after '$p$ $\rightarrow$ data' is updated but before '$p$ $\rightarrow$ state' is updated. In this case, the memory state in PM is still not correct for data recovery. To handle this issue, durable transactions \cite{ref6,ref8,ref26} (e.g. undo logging), which provide all-or-nothing guarantees by undoing changes from an aborted transaction, are required. To be able to roll back changes, an undo log entry will be created prior to every update performed within the transaction. The undo log entry contains the current value of the persistent structure that is to be updated. Once the log entry has been created and persisted, only then is the actual value updated. If a transaction succeeds, a commit message is atomically sent to invalidate the log entries belonging to the transaction. If a transaction fails, the recovery process uses all the persisted log entries to roll back partial changes from that transaction. 

\begin{figure}[htbp]
\centerline{\includegraphics{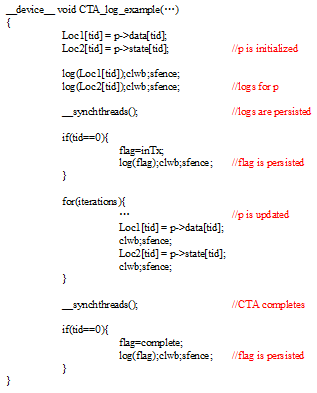}}
\caption{An example of CTA-level undo logging of a GPU kernel.}
\label{fig:background}
\end{figure}

Currently, memory persistency models, as well as durable transactions, are designed for the CPU. These models need to be re-architected for GPUs due to the differences in both the workloads and the processors' architectures. Lin et al. \cite{ref31} explore the implementation of different levels (kernel-level, CTA-level, and loop-level) of memory persistency models in GPU. We exploit their implementation, especially the CTA-level undo logging for GPU, in this study. Figure \ref{fig:background} shows an example of CTA-level undo logging. First, undo logs are created for the output elements (\textbf{Loc1} and \textbf{Loc2}) that are to be updated by the CTA. After ensuring all the threads persist their log using the $sfence$ followed by the CUDA $\_\_syncthreads$ function, a flag for the log is set to be \textbf{inTx} and is made durable. The values (\textbf{p$\rightarrow$data} and \textbf{p$\rightarrow$state}) of the output elements are updated iteratively by the CTA. At the end of the CTA, all the outputs are ensured to be persisted using another $\_\_syncthread$ function, and the flag is set to be \textbf{complete}. Note that all the undo logging described in the following sections in this study refers to CTA-level undo logging.

\subsection{GPU with PM}
Today, there is no hardware with PM onboard the GPU. If a GPU application wishes to leverage PM’s persistence, it would typically perform computations on the GPU and ensure the persistence of the results on the CPU. The computation results need to be transferred from the GPU's memory to the CPU’s memory and rely on the CPU to guarantee data recoverability. Alternatively, one could leverage the file system, since only the file systems atop block-storage devices could guarantee persistence before the advent of NVM technologies. After the computation results from the GPU are transferred to the CPU's memory, the CPU writes the results to a PM-resident file and then guarantees persistence using the $fsync$ function. Furthermore, a PM resident file can be memory-mapped onto the CPU’s address space, and the GPU's results can be transferred to the memory-mapped file residing in the CPU memory using $cudaMemcpy$.

Pandey et al. proposed GPM \cite{ref21} to use NVIDIA’s uniform virtual addressing (UVA) technique \cite{ref32} to map PM to GPU’s address space, and system-scoped fence with selective disabling of the data direct IO (DDIO) feature to enable GPUs to access and persist data in PM with no CPU involvement. GPM leverage the fact that the modern PM (Intel Optane \cite{ref14}) is placed alongside the DRAM, as in a typical Intel Xeon server, and can be accessed by a GPU over the PCI-e interconnect. In this case, UVA can be used to map desired portions of NVM onto the virtual address space of a GPU kernel. Then, GPU kernels can directly access and manipulate PM-resident data structures at byte granularity using $load$s and $store$s without the CPU's assistance. 

\section{Motivation}
\label{sec:moti}

\begin{figure}[htbp]
\centerline{\includegraphics[width=.4\textwidth]{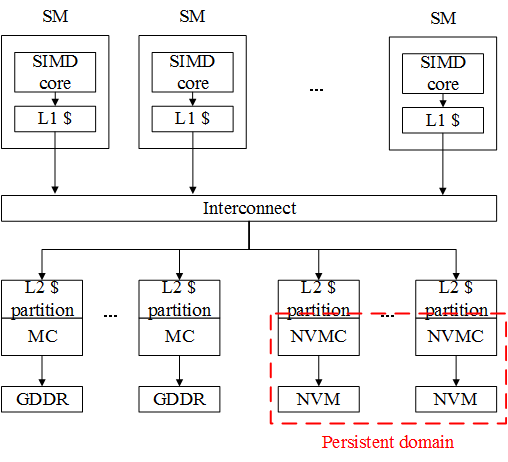}}
\caption{Overviews of the assumed GPU system equipped with NVM/PM. \$ denotes cache. MC denotes the memory controller. NVMC denotes the non-volatile memory controller.}
\label{fig:1}
\end{figure}

This section discussed the impetus of this study. For simplicity, we assume a discrete GPU system equipped with persistent memory (Figure \ref{fig:1}). This assumption enables us to focus on the GPU memory hierarchy without considering the potential effects of the host-side memory system as well as the costly data transfers between the system and GPU device memory. We further assume that the PM controller supports the asynchronous DRAM refresh (ADR) \cite{ref33} feature and is in the persistent domain. All updates to PM will be durable once they reach the PM controller. Intel recently announced an enhanced ADR (eADR) \cite{ref34} feature. eADR drains the entire contents of CPU caches to PM on power failures, which obviates the need to flush cache blocks from the CPU’s caches in order to guarantee persistence in future processors. Besides, the fence is still needed to maintain the ordering of writes to PM \cite{ref35}. This study can be projected on a future system with eADR, since the data path selection of data persisting takes place before the cache line flushes. We focus on GPUs' durable transactions (undo logging) in this study. Note that there are other persistency memory models for persist ordering in GPU, these models are beyond the scope of this study. 

\begin{table}[htbp]
\caption{Different types of benchmarks are categorized by the performance difference using the temporal and the non-temporal path for undo logging.}
\begin{center}
\begin{tabular}{|c|c|c|}
\hline
\textbf{Type} & \textbf{\textit{Metric}}& \textbf{\textit{Benchmark}}\\
\hline
\uppercase\expandafter{\romannumeral1}& ${perf}^T$ $>$ ${perf}^{NT}$ , $diff$ $>$ $5\%$ & SAD1, GRID1, GRID2,\\
 &  & 2DCONV, Backprop1,\\
 &  & Pathfinder, 2MM1,\\
 &  & 3MM1, GEMM\\
\hline
\uppercase\expandafter{\romannumeral2}& ${perf}^T$ $<$ ${perf}^{NT}$ , $diff$ $>$ $5\%$ & SGEMM, Backprop2,\\
 &  & ATAX1, ATAX2, NW,\\
 &  & SSSP2, MVT1,\\
 &  & GESUMMV, RA\\
\hline
\uppercase\expandafter{\romannumeral3}& $diff$ $\leq$ $5\%$ & Stencil1, SSSP1,\\
&  & StreamTriad, BFS\\
\hline
\end{tabular}
\label{tab:broadType}
\end{center}
\end{table}

\begin{figure*}[htbp]
\centerline{\includegraphics[width=.9\textwidth]{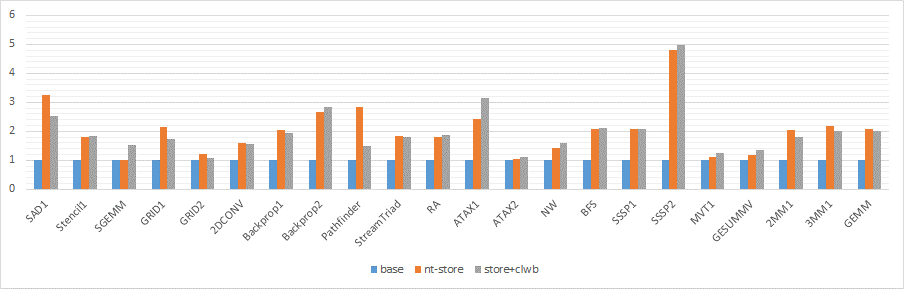}}
\caption{Normalized execution time of 22 GPU benchmark kernels using separate data paths for undo-logging. ATAX1 and ATAX2 indicate the first and the second kernel of the ATAX benchmark. This goes the same for the other benchmarks. $base$ denotes the benchmarks that are executed using NVM without persistency support. $nt$-$store$ denotes the implementation follows the description in \cite{ref27,ref28}, which uses the non-temporal path to persist the log updates to PM and uses the temporal path for the subsequent data updates. $store$+$clwb$ denotes that both the log and the corresponding data updates use the temporal path. All the results are normalized by the execution time running with $base$.}
\label{fig:moti}
\end{figure*}

Figure \ref{fig:moti} shows the experimental results of 22 GPU benchmark kernels using a static data path selection for undo-logging. 
Figure \ref{fig:moti} reveals several important observations. \textbf{First, data recoverability enabled by undo logging introduces different levels of performance overheads to different GPU benchmark kernels.} For instance, the PM operations invoking the $nt$-$store$ instructions expand the execution time of the SAD1 kernel to 325.5\% compared to the execution time without undo logging, while $nt$-$store$ expands SGEMM's execution time to 102.7\%. 
Furthermore, we can see that $store$+$clwb$ delivers a shorter execution time than $nt$-$store$ among several benchmark kernels (SAD1, GRID1, GRID2, 2DCONV, Pathfinder, 2MM1, 3MM1, GEMM). These results lead to our second observation: \textbf{the presumption that $nt$-$store$ is more beneficial for undo logging is not applicable to all GPU applications.} According to Figure \ref{fig:moti}, we broadly categorized 22 benchmark kernels into three types according to two metrics ($perf$ and $diff$) as described in Table \ref{tab:broadType}. $perf^T$ denotes the benchmark performance (execution time, etc.) using the temporal path ($store$+$clwb$) for undo logging, and $perf^{NT}$ denotes the performance using the non-temporal path ($nt$-$store$). $diff$ denotes the absolute difference between $perf^T$ and $perf^{NT}$. 

\begin{figure}[htbp]
\centerline{\includegraphics[width=.4\textwidth]{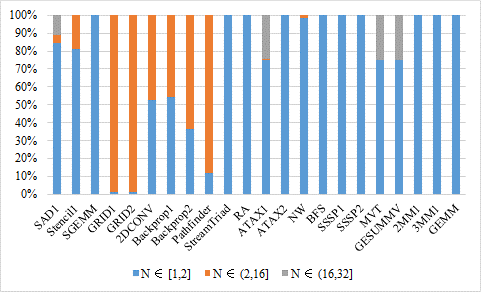}}
\caption{Distribution of $load$s/$store$s with different memory un-coalescing degrees. $N$ denotes the number of memory accesses for a $load$/$store$ belonging to the same warp after memory coalescing.}
\label{fig:moti_distrib}
\end{figure}

Due to the similarity between this study and the cache bypassing problem \cite{ref36,ref37} in GPU, we try to explain the benchmark kernels' performance difference following the cache bypassing logic. One of the most important ideas in this area is to perform cache bypassing according to GPU kernels' memory coalescing behaviors \cite{ref36}. These studies conclude that the un-coalesced $load$s/$store$s should bypass the volatile caches because such requests generate massive memory accesses compared to the other memory instructions. Besides, data fetched by these requests is likely to possess poor locality, which is not worthy of consuming the scarce cache resources in the GPU architecture. Figure \ref{fig:moti_distrib} shows the distribution of the $load$s/$store$s (including the undo loggings and the other memory updates) with different numbers of memory accesses. We can see that the performance difference shown in Figure \ref{fig:moti} is not coherent with the benchmark kernels' coalescing behaviors. For example, StreamTriad, ATAX2, and 3MM1 are perfectly memory coalesced, whose memory un-coalescing degrees are within $[1, 2]$. However, these kernels vary from type \uppercase\expandafter{\romannumeral1} to \uppercase\expandafter{\romannumeral3} as described in Table \ref{tab:broadType}. Thus, we can see that this study is not essentially identical to the cache bypassing problem in GPU, and a new design is required.

\begin{table*}[htbp]
\caption{Temporal and spatial locality of 22 GPU benchmark kernels' memory accesses in GPU's L1D and L2 cache. $l1d$ denotes the L1D cache. $l2$ denotes the L2 cache. $t$ denotes temporal. $s$ denotes spatial. $all$ denotes the memory access generated by both the log and the other data updates. $log$ denotes the memory access generated by the log updates. $type$ denotes the kernels' type as described in Table \ref{tab:broadType}. $reason$ denotes the reason why the certain kernel belongs to a certain type as described in Table \ref{tab:reason}.}
\begin{center}
\begin{tabular}{|c|c|c|c|c|c|c|c|c|c|c|c|c|}
\hline
\textbf{Benchmark}& 
\textbf{\textit{l1d\_t\_all}}& \textbf{\textit{l1d\_t\_log}}& \textbf{\textit{l1d\_s\_all}}& \textbf{\textit{l1d\_s\_log}}& \textbf{\textit{l2\_t\_all}}& \textbf{\textit{l2\_t\_log}}& \textbf{\textit{l2\_s\_all}}& \textbf{\textit{l2\_s\_log}}& \textbf{\textit{all}}& \textbf{\textit{log}}& \textbf{\textit{type}}& \textbf{\textit{reason}}\\
\hline
SAD1 & 0 
 & 0 & 0
 & 0 & 289160
 & 157652 & 300509
 & 167001 & 2288857
 & 324803 & \uppercase\expandafter{\romannumeral1}
 & c\\
\hline
Stencil1 & 0
 & 0 & 0
 & 0 & 0
 & 0 & 0
 & 0 & 131871
 & 33849 & \uppercase\expandafter{\romannumeral3}
 & f\\
\hline
SGEMM & 0
 & 0 & 0
 & 0 & 0
 & 0 & 0
 & 0 & 3191435
 & 133274 & \uppercase\expandafter{\romannumeral2}
 & h\\
\hline
GRID1 & 2398
 & 0 & 3126
 & 0 & 7495
 & 3483 & 9588
 & 3930 & 1047435
 & 662069 & \uppercase\expandafter{\romannumeral1}
 & a\\
\hline
GRID2 & 47898
 & 0 & 47898
 & 0 & 140745
 & 64156 & 140745
 & 64156 & 591011
 & 200163 & \uppercase\expandafter{\romannumeral1}
 & a\\
\hline
2DCONV & 0
 & 0 & 0
 & 0 & 0
 & 0 & 0
 & 0 & 79334
 & 17060 & \uppercase\expandafter{\romannumeral3}
 & f\\
\hline
Backprop1 & 0
 & 0 & 7
 & 6 & 0
 & 0 & 153
 & 147 & 88412
 & 65076 & \uppercase\expandafter{\romannumeral1}
 & g\\
\hline
Backprop2 & 8764
 & 2671 & 9476
 & 2987 & 8958
 & 6330 & 9707
 & 6920 & 280763
 & 118159 & \uppercase\expandafter{\romannumeral2}
 & b\\
\hline
Pathfinder & 1221
 & 97 & 1225
 & 101 & 35000
 & 26646 & 35843
 & 27229 & 806640
 & 382582 & \uppercase\expandafter{\romannumeral1}
 & d\\
\hline
StreamTriad & 0
 & 0 & 0
 & 0 & 0
 & 0 & 0
 & 0 & 36000
 & 17476 & \uppercase\expandafter{\romannumeral3}
 & f\\
\hline
RA & 1
 & 1 & 2
 & 2 & 2
 & 1 & 4
 & 2 & 6644
 & 2676 & \uppercase\expandafter{\romannumeral2}
 & b\\
\hline
ATAX1 & 45460
 & 45460 & 45460
 & 45460 & 135526
 & 89964 & 135526
 & 89964 & 1926411
 & 180170 & \uppercase\expandafter{\romannumeral2}
 & b\\
\hline
ATAX2 & 89032
 & 45562 & 134032
 & 45562 & 135770
 & 90000 & 135770
 & 90000 & 545022
 & 180170 & \uppercase\expandafter{\romannumeral2}
 & b\\
\hline
NW & 82
 & 40 & 154
 & 84 & 204
 & 88 & 426
 & 156 & 3195
 & 1392 & \uppercase\expandafter{\romannumeral2}
 & h\\
\hline
BFS & 5
 & 0 & 5
 & 0 & 2
 & 1 & 2
 & 1 & 9528
 & 79 & \uppercase\expandafter{\romannumeral3}
 & e\\
\hline
SSSP1 & 0
 & 0 & 0
 & 0 & 1
 & 0 & 1
 & 0 & 1007
 & 17 & \uppercase\expandafter{\romannumeral3}
 & e\\
\hline
SSSP2 & 1028
 & 771 & 1028
 & 771 & 2570
 & 1285 & 2570
 & 1285 & 9509
 & 6168 & \uppercase\expandafter{\romannumeral2}
 & b\\
\hline
MVT & 63720
 & 63720 & 63720
 & 63720 & 196227
 & 130846 & 196227
 & 130846 & 9444863
 & 341037 & \uppercase\expandafter{\romannumeral2}
 & b\\
\hline
GESUMMV & 60929
 & 60929 & 60929
 & 60929 & 193639
 & 129104 & 193639
 & 129104 & 15752038
 & 281725 & \uppercase\expandafter{\romannumeral2}
 & b\\
\hline
2MM1 & 1639464
 & 0 & 2809772
 & 0 & 5608094
 & 4905025 & 6369854
 & 5416431 & 95599514
 & 51796409 & \uppercase\expandafter{\romannumeral1}
 & a\\
\hline
3MM1 & 28360
 & 0 & 28608
 & 0 & 441916
 & 411072 & 449205
 & 418361 & 7281497
 & 4458165 & \uppercase\expandafter{\romannumeral1}
 & a\\
\hline
GEMM & 33545
 & 2031 & 34045
 & 2031 & 1533475
 & 1501161 & 1559512
 & 1527197 & 7332760
 & 4503201 & \uppercase\expandafter{\romannumeral1}
 & d\\
\hline
\end{tabular}
\label{tab:charac}
\end{center}
\end{table*}

In order to explain the performance difference using different data paths, we define 10 variables to characterize the GPU kernels' temporal and spatial localities as described in Table \ref{tab:charac}. For example, $l1d\_t\_all$ denotes the number of cache hits with the temporal locality in the L1D cache, and these accesses are generated by both the log and the other data updates. This goes the same for the other 7 variables ($l1d(l2)\_t(s)\_all(log)$). Since we are focusing on durable transactions in GPU PM, we believe that the statistics of both the log updates ($log$) and the non-log updates ($all$) are required. We define that the non-log updates' temporal locality in the L1D cache means the re-hits of a specific byte by a log update followed by a non-log data update, and the log updates' temporal locality in the L1D cache means the re-hits of a specific byte by a log update followed by another log update. Similarly, we define that the non-log updates' spatial locality in the L1D cache means the re-hits of a specific aligned 128-byte segment (cache block) by a log update followed by a non-log data update, and the log updates' spatial locality in the L1D cache means the re-hits of a specific byte by a log update followed by another log update. These go the same for the statistics of the L2 cache. Note that Maxwell, Pascal, and Volta GPU architectures use demand-fetch caches to fetch only the chunks that are requested instead of fetching a full cache line to handle the data over-fetch problem \cite{ref38}. Thus, the number of accesses in the GPU caches with the temporal locality is not necessarily identical to the one with the spatial locality. According to Table \ref{tab:charac}, we conclude 8 reasons to explain Figure \ref{fig:moti}'s performance difference as described in Table \ref{tab:reason}.

\begin{table}[htbp]
\caption{Statistics-based reasoning.}
\begin{center}
\begin{tabular}{|l|l|l|}
\hline
\textbf{Reason} & \textbf{\textit{Metrics}}& \textbf{\textit{Type}}\\
\hline
a & $l1d\_t\_all$ $\neq$ $0$ $and$ $l1d\_s\_all$ $\neq$ $0$ $and$ & \uppercase\expandafter{\romannumeral1}\\
&  $l1d\_t\_log$ $=$ $0$ $and$ $l1d\_s\_log$ $=$ $0$  & \\
\hline
b & $l1d\_t\_all$ $\neq$ $0$ $and$ $l1d\_s\_all$ $\neq$ $0$ $and$ & \uppercase\expandafter{\romannumeral2}\\
& $l1d\_t\_log$ $\neq$ $0$ $and$ $l1d\_s\_log$ $\neq$ $0$ $and$ & \\
& $l1d\_t(s)\_log$ / $l1d\_t(s)\_all$ $>$ 0.25 & \\
\hline
c & $l1d\_t\_all$ $=$ $0$ $and$ $l1d\_s\_all$ $=$ $0$ $and$ & \uppercase\expandafter{\romannumeral1}\\
& $l1d\_t\_log$ $=$ $0$ $and$ $l1d\_s\_log$ $=$ $0$ $and$ & \\
& $l2\_t\_all$ $\neq$ $0$ $and$ $l2\_s\_all$ $\neq$ $0$ $and$ & \\
& $l2\_t\_log$ $\neq$ $0$ $and$ $l2\_s\_log$ $\neq$ $0$ & \\
\hline
d & $l1d\_t\_all$ $\neq$ $0$ $and$ $l1d\_s\_all$ $\neq$ $0$ $and$ & \uppercase\expandafter{\romannumeral1}\\
& $l1d\_t\_log$ $\neq$ $0$ $and$ $l1d\_s\_log$ $\neq$ $0$ $and$ & \\
& $l1d\_t(s)\_log$ / $l1d\_t(s)\_all$ $\leq$ 0.25 & \\
\hline
e & $log$ $\le$ $100$ & \uppercase\expandafter{\romannumeral3}\\
\hline
f & $l1d\_t\_all$ $=$ $0$ $and$ $l1d\_s\_all$ $=$ $0$ $and$ & \uppercase\expandafter{\romannumeral2}, \uppercase\expandafter{\romannumeral3}\\
& $l1d\_t\_log$ $=$ $0$ $and$ $l1d\_s\_log$ $=$ $0$ $and$ & \\
& $l2\_t\_all$ $=$ $0$ $and$ $l2\_s\_all$ $=$ $0$ $and$ & \\
& $l2\_t\_log$ $=$ $0$ $and$ $l2\_s\_log$ $=$ $0$ & \\
\hline
g & $l1d\_t\_all$ $=$ $0$ $and$ $l1d\_s\_all$ $\neq$ $0$ $and$ & \uppercase\expandafter{\romannumeral1}\\
& $l1d\_t\_log$ $=$ $0$ $and$ $l1d\_s\_log$ $\neq$ $0$ & \\
\hline
h & usage of shared memory & \uppercase\expandafter{\romannumeral2}\\
\hline
\end{tabular}
\label{tab:reason}
\end{center}
\end{table}

\begin{figure}[htbp]
\centerline{\includegraphics[width=.4\textwidth]{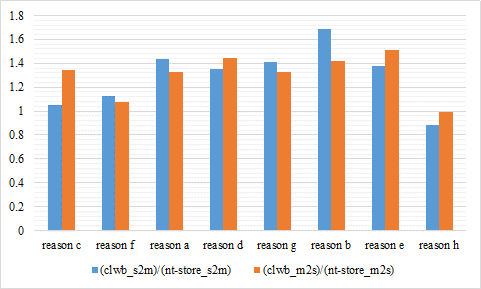}}
\caption{Normalized mean transmitted bytes comparison between the GPU L1D cache and the L2 cache using the $store$+$clwb$ and the $nt$-$store$ instructions. $s2m$ denotes the transmission from the L1D cache to the L2 cache, and $m2s$ denotes the transmission from the L2 cache to the L1D cache.}
\label{fig:moti_HR}
\end{figure}

Intuitively, the $nt$-$store$ instruction is considered profitable for log updates \cite{ref29,ref13}. This is because logs are considered only necessary for crash consistency and never read in failure-free execution. In this case, the $nt$-$store$'s write-no-allocate policy, which writes data blocks directly to memory without adding to the cache, provides more benefits for logging and reduces cache pollution. On the other hand, introducing the frequently reused data blocks into the volatile caches by the $store$+$clwb$'s write-allocate policy help save a lot of expensive memory transactions. Thus, a simple idea for the undo logging's data path selection is to rely on the locality between the log updates and the subsequent data updates. More specifically, the stronger locality is observed the more likely to use the temporal path ($store$+$clwb$). Otherwise, the non-temporal path ($nt$-$store$) should be selected. However, Table \ref{tab:charac} shows different patterns compared to this intuition. And this leads to our third observation: \textbf{$store$+$clwb$ do not always provide more benefits than $nt$-$store$ for data updates with good localities in GPU caches}. For reasons $a$, $b$, and $d$, the major differences come from whether the values of $l1d\_t\_log$ and $l1d\_s\_log$ are equal to 0 or not. $l1d\_t\_log$ and $l1d\_s\_log$ reveal whether the identical cache chunk or the identical cache line is repeatedly referenced by the log updates. When these two variables are equal to 0 (reason $a$) or only occupy a small fraction ($\leq$ 0.25) of all the memory accesses with the locality (reason $d$), the experimental results follow the aforementioned intuition that using the $store$+$clwb$ instructions achieve better performance than using the $nt$-$store$ (type \uppercase\expandafter{\romannumeral1}). However, when too-strong localities are observed in the L1D cache accesses caused by the log updates (reason $b$), we can see that $nt$-$store$ provides more benefits compared to $store$+$clwb$ in this case. This is because the cache thrashing between the L1D cache and the L2 cache through the slow interconnected network impedes the performance advantage of using the caches for the data with good localities. 
Figure \ref{fig:moti_HR} shows the normalized mean transmitted bytes of the kernels belonging to a certain reason using the temporal ($store$+$clwb$) and the non-temporal ($nt$-$store$) data path. 
When the data localities in volatile caches increase, the memory traffic as well as the pipeline utilization grow due to the less warp stall and the fewer expensive memory transactions. However, when such localities become too strong and the reuse distances are too long to help the cached data be reused before eviction, the invalidation frequency of the GPU volatile caches will grow dramatically which finally turns into cache contention and thrashing. 
For kernels belonging to reason $c$, since no locality is observed in the L1D cache, there is no risk of experiencing serious thrashing while enjoying the benefit provided by $store$+$clwb$, these kernels are categorized as type \uppercase\expandafter{\romannumeral1}. For reason $f$, kernels have no data locality in either the L1D cache or the L2 cache. $nt$-$store$ is intuitively considered suitable for this pattern because of the write-no-allocate policy. However, these GPU kernels show interesting behavior in that the performance of using the $store$+$clwb$ and the $nt$-$store$ are similar (type \uppercase\expandafter{\romannumeral3}). We believe that this is because of GPU's massive multi-threading which effectively hides the cache allocation overhead, and this results in average performance using either the temporal path or the non-temporal path for persistency. For reason $g$, moderate localities are observed in both the L1D cache and the L2 cache, thus kernels belonging to this category are type \uppercase\expandafter{\romannumeral1}. For reason $e$ and reason $h$, these are two special cases that can not be explained by the adaptability between the log and the non-log updates' localities and the temporal or the non-temporal data path. For reason $e$, the number of the log updates is too small ($<$100) and this makes a negligible difference between the usage of the $store$+$clwb$ and the $nt$-$store$ (type \uppercase\expandafter{\romannumeral3}). For reason $h$, these kernels use the shared memory to communicate and synchronize for the threads within the same CTA(TB). In this case, most of the data fetched in the volatile cache will not be reused, and using the $nt$-$store$ instruction becomes more beneficial than using the $store$+$clwb$ instructions (type \uppercase\expandafter{\romannumeral2}).

\section{Our solution}
\label{sec:sol}
As discussed in Section \ref{sec:moti}, we leverage 10 variables and 8 reasons to explain the performance difference of GPU' undo logging using the $store$+$clwb$ and the $nt$-$store$ instructions. Based on these observations, we propose adaptive data path selection for durable transactions in GPU PM (AGPM). The key idea is to alter the log updates' data path adaptively according to statistics tracking the localities between the log and the non-log updates within the GPU L1D and the L2 caches. The temporal data path ($store$+$clwb$) should be selected for log updates when moderate localities are presented between the L1D and the L2 cache, or there are no risks of contention among different levels of volatile caches through the GPU interconnected network. The non-temporal path should be selected for log updates when too-strong localities are simultaneously presented in more than one GPU cache. For cases when no localities are presented inside the GPU memory hierarchy, the selection of the path is flexible and the non-temporal path is recommended to use to reserve the scarce cache resources for other data with high localities. For special cases when the shared memory is used or the number of log updates is too small, the non-temporal path is recommended to avoid wasting the volatile caches.

\subsection{Period-based strategy}
\label{sec:AGPM_period}
Although we want to leverage observations described in Section \ref{sec:moti} to achieve adaptive data path selection, results in Table \ref{tab:charac} are not amenable for direct use in runtime execution since they are recorded when the kernels are completed. In order to enable undo logging's adaptive path selection by capturing the patterns revealed in Table \ref{tab:charac}, we collect statistics in periods during kernel execution and we use these statistics to feed a path selector as shown in Figure \ref{fig:arch_AGPM}. This is further discussed in Section \ref{sec:AGPM_arch}.

It is challenging to determine a cycle-based period for all types of GPU kernels. This is because the execution time of different GPU kernels or different CTAs belonging to the same kernel may vary vastly, and this leads to different levels (kernel-level, CTA-level, iteration-level, etc.) of memory persistency models in GPU \cite{ref31}. 
In this study, a period is defined as a thresholded number of log updates to PM. When the number of logs exceeds the threshold, all the AGPM buffers will be flushed and data localities will be re-captured in the new period. According to Table \ref{tab:charac}, we initialize this threshold as 10000. We define a metric, cycles-waited-per-PM-request (CWPPR), to update the threshold. CWPPR denotes the average number of cycles that each PM request finishes servicing. We exploit this metric and apply a sampling method that updates the threshold every time one period is completed. We define the current period ($P_1$) as the sampling period. If the CWPPR of the sampling period ($P_1$) is bigger than the previous period ($P_0$), we consider the length of the threshold is not long enough to capture the right pattern. In this case, the threshold is increased by 10\% of the previous threshold. For instance, if the previous threshold is 10000, then the updated threshold is 11000. Otherwise, the threshold is decreased by 10\% of the previous threshold. Finally, the updated threshold is stored and used in the next sampling period. In some cases when the number of log updates to PM of the entire GPU kernel is less than 10000, the threshold will be initialized with a smaller value.

\subsection{AGPM architecture}
\label{sec:AGPM_arch}
\begin{figure*}[htbp]
\centerline{\includegraphics[width=.9\textwidth]{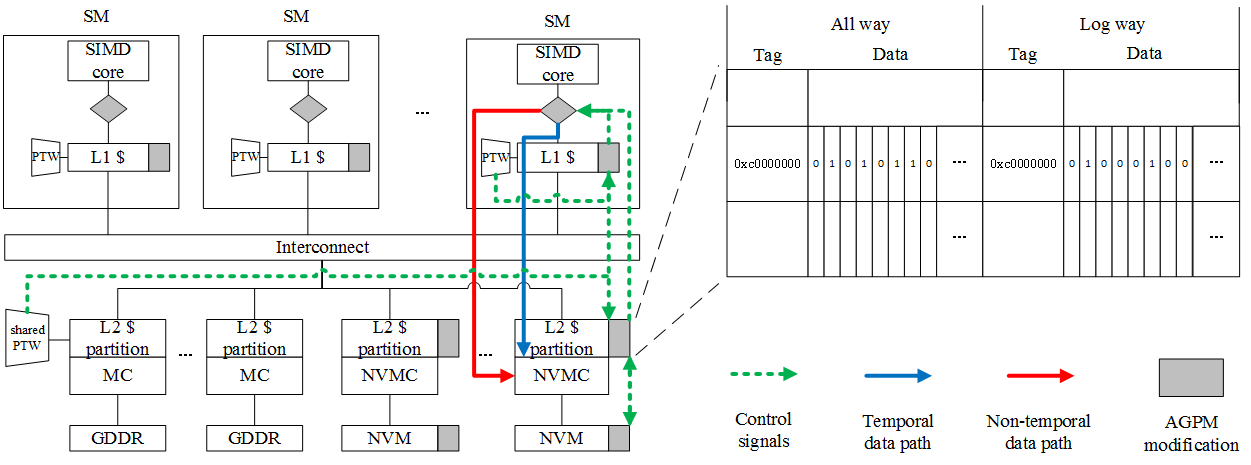}}
\caption{Overview of AGPM's architecture.}
\label{fig:arch_AGPM}
\end{figure*}

Figure \ref{fig:arch_AGPM} shows the overall architecture of AGPM. The gray rhombus between the SIMD core and the L1 cache inside the GPU SM denotes the path selector which determines the memory requests from a PM log updates to use a specific data path to reach the persistent domain (NVMC in this study). The gray blocks attached to the L1 cache and the L2 cache partition denote the AGPM buffers which are used to capture the localities between the log updates and the subsequent data updates. The gray blocks attached to the non-volatile memory denote the reservation buffers which are used to keep the information flushed from the AGPM buffer. The blue and the red solid lines denote the temporal and the non-temporal data path respectively. The green dashed line denotes the control signals to instruct the selector which path should be selected for undo logging. Note that Figure \ref{fig:arch_AGPM} demonstrates the AGPM-related memory transactions between one SM and one memory partition, and these transactions can be performed among different SMs and different non-volatile memory partitions in the GPU.

After coalescing in LD/ST, PM requests from a memory instruction may need to access the L1D cache. In AGPM, these requests are first sent to a path selector. According to the requested addresses and the corresponding statistics recorded in the AGPM buffers extended in the L1D and the L2 cache, if patterns that are identical to the reason $a$, $c$, $d$, and $g$ as described in Table \ref{tab:reason} are observed, the requests from the same warp which execute the PM instructions will access the GPU's volatile caches. Since ADR is assumed to be supported in the non-volatile memory controller (NVMC), data is persisted when the dirty cache lines are sent to the NVMC from the L2 cache partition. Otherwise, if patterns that are identical to the reason $b$, $e$, $f$, $g$, and $h$ are observed, the requests will bypass the caches and they will be directly inserted into the write-pending queue (WPQ) residing in the NVMC. In this case, data becomes persisted once they reach NVMC without taking extra actions.

The structure of the AGPM buffer is similar to a 2-way set-associative cache (depicted in Figure \ref{fig:arch_AGPM}). One way is for the statistics of the exclusive PM log updates ($log$ way), and the other way is for the statistics of all kinds of memory requests ($all$ way). In each way, every entry in the set-associative cache has two fields: tag and data. The tag is an aligned 128-byte cache block address and the data is a counter vector, which records the number of references to each byte in a cache block. Besides the byte counters, two extra counters are used in the data field. One extra counter is used to record the number of references to the specific cache block. Another extra counter is used as a mark for the change of the data path. If a certain log update's data path is changed from the temporal one to the non-temporal one, then the original $clwb$ should be dropped since the data is sent directly to the persistent memory controller. Otherwise, if the data path is changed from the non-temporal one to the temporal one, then a $clwb$-like operation needs to be added after the temporal $store$ to guarantee the data persistence. When a new PM request is made, an associated entry will be initialized with 0 and added in both ways in the AGPM buffer. The counters of the requested bytes are increased by 1. Then, for a subsequent memory request (both the PM requests and the others), the page table walker (PTW) begins a walk. Once the walker knows the request is a hit, it notifies the AGPM buffer with the byte address. Then the cache block address and the offset are calculated. The AGPM buffer searches the entries with the cache block address. If there is a miss in the AGPM buffer and the request is not from a PM instruction, then no actions will be taken. If there is a hit, according to the offset, if the corresponding byte counter is 0, then only the cache block counter will be increased. If the byte counter is larger than 0, then both the byte and the cache block counters will be increased. 
If the subsequent request belongs to a PM instruction, counters in both ways of the AGPM buffer will be updated. Otherwise, only the counters in the $all$ way will be updated. When the AGPM buffers are enquired for path selection, variables described in Table \ref{tab:reason} are calculated using the recorded counters. For example, $l1d\_t\_all$ is calculated as the sum of all the byte counters whose values are larger than 1 in the $all$ way of the AGPM buffers at the L1 cache level, and $l1d\_s\_all$ is calculated as the sum of both the byte counters  whose values are larger than 1 and the cache block counters in the $all$ way of the AGPM buffers at the L1 cache level. These go the same for the other variables described in Table \ref{tab:reason}.

When a cache block is flushed to a lower memory hierarchy, the associated AGPM entry will also be flushed to a lower-level AGPM buffer. When a cache block recorded in the AGPM buffer is eventually written back either due to a regular replacement or due to a $clwb$-like instruction, the associated content in the AGPM buffer will be first placed in the reservation buffer residing in the GPU memory, and then flushed from the AGPM buffer. Once this cache block is fetched in the volatile caches again before the current period or the kernel is completed, the reserved information in the GPU memory will be re-filled into the AGPM buffer. When a period ends or the kernel execution completes, all the contents in the AGPM buffer and the reservation buffer will be flushed.


\subsection{Overheads analysis}
We used a 2-way AGPM buffer with 1024 entries to record the localities between the log updates and the subsequent data updates. Each entry has a tag and a data field. Assume the system is 64-bit wide, then the tag is 57 bits. The data has 128+2=130 counters, and a 5-bit counter can meet the requirement in most cases. Thus, the data field needs 650 bits in total. An entry needs 707 bits (about 89 B), and 1024 entries translate to 89 KB storage cost. As the memory access patterns are nearly the same among different SMs, optimistically only 1 AGPM buffer is needed to be implemented to track the L1-level data localities. Another AGPM buffer is required to be shared among all the L2 partitions. The total cost will be 89*2=178 KB.

\begin{figure*}[htbp]
\centerline{\includegraphics[width=.9\textwidth]{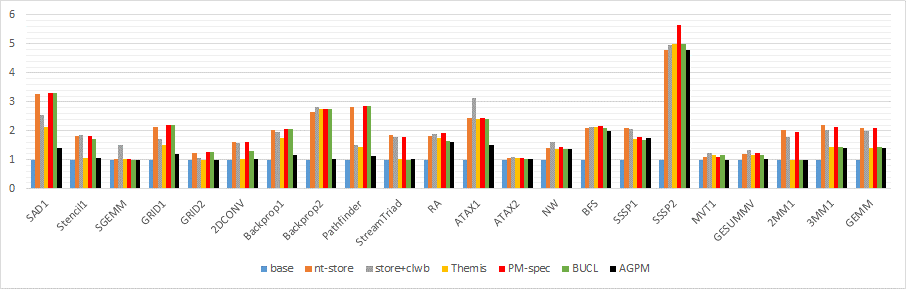}}
\caption{Normalized execution time of 22 GPU benchmark kernels using different strategies for undo-logging's data path selection. All the results are normalized by the execution time running with $base$.}
\label{fig:eval_perf1}
\end{figure*}

\section{Evaluation}
\label{sec:eval}
We evaluate AGPM by comparing it against the conventional memory persistency models which use a static path strategy for undo logging. We further compare the benchmarks' normalized execution time using AGPM against using the other state-of-the-art multi-data-path architectures to PM and cache bypassing methods to verify the proposed solution's feasibility for the GPU PM applications. 

\subsection{Evaluation methodology}
We use a GPGPU-Sim extension implemented by \cite{ref31}. This extension provides functional and timing simulation support for memory persistency models which are re-architected for GPUs. These models are adapted and optimized according to GPU workloads' characteristics, GPU's bandwidth-sensitive feature, and GPU's memory hierarchy. To leverage this extension, the compiler uses inline assembly to insert the PM instructions such as $clwb$ and $sfence$, and the simulator is modified to support the semantics of these instructions. In our experiments, durable transactions are supported by software-based undo logging. More specifically, before a transaction starts, an undo log is created by making a copy of the data to be updated and this log is persisted. A flag is set and persisted to indicate that the transaction is in process. During this transaction, data are updated and persisted. The flag is updated when the transaction completes, and the undo log will be released. With undo logging, the recovery code checks the flag to find out the status of a transaction. If the transaction is interrupted before its completion, the undo log will be applied to restore the data.

Our experiments use a set of regular and irregular applications from Rodinia \cite{ref39}, Lonestar \cite{ref41}, Polybench \cite{ref40}, and Parboil \cite{ref42} GPU benchmark suites. These benchmarks are modified to use the PM instructions to construct durable transactions, which are similar to the implementation of \cite{ref31}. The simulation configurations of GPGPU-Sim are shown in Table \ref{tab:config}.

\begin{table}[htbp]
\caption{Configuration parameters of GPGPU-Sim.}
\begin{center}
\begin{tabular}{|l|l|}
\hline
\textbf{GPU cores} & 20 SMs, SIMD width=32, 1.8GHz\\
\hline
\textbf{Shader Core Config} & Max 2048 threads and 64 warps and 32 CTAs\\
 & per SM, 32 threads per warp, 4 GTO scheduler\\
\hline
\textbf{Per-SM L1D-cache} & 24KB, 128B line, 6-way associativity,\\
 & 256 MSHRs\\
\hline
\textbf{Per-SM SMEM} & 96KB, 32 banks\\
\hline
\textbf{Shared L2 cache} & 2048KB, 128KB/partition, 128B line,\\
 & 16-way associativity, 256 MSHRs\\
\hline
\textbf{L1D/L2 policies} & XOR-indexing, allocate-on-miss, LRU,\\
 & L1D:WEWN, L2:WBWA\\
\hline
\textbf{Interconnect} & 16*16 crossbar, 32B flit size, 1.4GHz\\
\hline
\textbf{Memory Controller} & 8 channels, 2 L2 banks/channel,\\
 & FR-FCFS scheduler, 1.2GHz, BW:307GB/s\\
\hline
\textbf{NVM latency} & Read: 160ns, Write: 480ns\\
\hline
\textbf{DRAM latency} & Read: 160ns, Write: 160ns\\
\hline
\end{tabular}
\label{tab:config}
\end{center}
\end{table}

\subsection{Performance comparison}
Figure \ref{fig:eval_perf1} shows the performance comparison of 22 GPU benchmark kernels using different data path selection strategies for undo logging. $nt$-$store$ and $store$+$clwb$ indicate that using a static strategy of using the non-temporal and the temporal data path for all the log updates. Themis \cite{ref27} presents a multi-data-path architecture to PM that differentiates temporal and non-temporal stores. Themis uses a non-temporal store path as a fast store path to PM, while temporal stores use a slow data path. Due to paths' latency difference, Themis can eliminate almost all persist-barriers, leading to higher performance of persistent applications. PM-spec \cite{ref28} also presents a multi-data-path architecture to PM. In contrast to Themis, PM-spec allows the PM controller to receive PM load and store with separate data paths respectively. Specifically, PM loads go to the temporal data path while PM stores through the non-temporal data path. BUCL \cite{ref36} is a GPU cache bypassing scheme for un-coalesced loads. If the number of un-coalesced memory access is bigger than the threshold of coalescing degree, the non-temporal data path will be selected. Otherwise, the temporal data path will be used. 

According to Figure \ref{fig:eval_perf1}, the average normalized execution time of using different strategies for undo logging is $nt$-$store$: 203.39\%, $store$+$clwb$: 198.43\%, Themis: 165.22\%, PM-spec: 207.29\%, BUCL: 185.60\%, and AGPM: 140.85\%. We can see that PM-spec achieves the longest normalized execution time even compared to the experimental results of the static strategies ($nt$-$store$, $store$+$clwb$). This is because the majority of the memory accesses generated from the PM log updates are $store$ operations. In this case, separating the data paths according to $load$ and $store$ is not suitable for undo logging. And this leads to the result that the performance of using PM-spec (207.29\%) is close to $nt$-$store$ (203.39\%). Except for the static strategies and PM-spec, BUCL achieves the second-longest normalized execution time. One of the reasons is that the bypassing logic of BUCL mainly relies on the coalescing degree of the GPU workloads. According to Figure \ref{fig:moti_distrib} and Table \ref{tab:broadType}, we can see that the GPU workloads' coalescing degrees do not always match their corresponding type. For example, according to BUCL's criteria and the coalescing degrees shown in Figure \ref{fig:moti_distrib}, SAD1, Stencil1, GRID1, GRID2, 2DCONV, Backprop1, Backprop2, Pathfinder, ATAX1, MVT1, and GESUMMV belong to the memory un-coalesced applications. However, SAD1, GRID1, GRID2, Backprop1, and Pathfinder belong type \uppercase\expandafter{\romannumeral1}. Stencil1 and 2DCONV belong to type \uppercase\expandafter{\romannumeral2}. Backprop2, ATAX1, MVT1, and GESUMMV belong to type \uppercase\expandafter{\romannumeral3}. In this case, using BUCL for data path selection can not provide equal performance improvement to PM log updates when it serves for GPU cache bypassing. Besides, BUCL is not designed for PM applications and this may lead to its incompatibility with undo logging transaction routine that a log update reaches persistence before the corresponding data updates. Whereas, BUCL delivers the best performance for some of the benchmarks (StreamTriad, SSSP1). This is because BUCL sends a memory request to the cache bypassing logic when there is a reservation fail after this request accesses the volatile caches. This feature prevents the interconnect network and the GPU memory hierarchy from being underutilized due to congested or stalled caches. Themis achieves the second-best performance among all the strategies. Themis delivers data path selection decisions relying on $store$'s temporal locality, and this is similar to AGPM. In this case, Themis achieves a comparatively shorter normalized execution time for most of the benchmarks except for Backprop2, ATAX1, ATAX2, MVT1, and GESUMMV. These benchmarks belong to reason $b$ described in Table \ref{tab:reason}, and Themis does not aware of the potential cache thrashing due to the too-strong data locality. In this case, temporal and non-temporal data paths are not differentiated and this leads to negatively affected performance using Themis. Note that one of the key advantages provided by Themis is to eliminate almost all persist-barriers of the persistency model. This feature achieves significant performance improvement for CPU workloads because the $sfence$ instruction precludes buffering, reordering of memory accesses, and out-of-order execution due to its ordering semantics. However, in the case of GPU where threads are highly overlapped and their memory latency could be hidden through massive multi-threading, the impact of removing the persist-barriers in GPU kernels is not as significant as it does in CPU workloads (1.5\% on average \cite{ref31}).


\section{Related Works}
\label{sec:related}
Intel provides PMDK \cite{ref8}, a library to leverage Optane NVM \cite{ref14}. Haria et al. \cite{ref17} propose a C++ library to allow multiple updates to one or more data structures to be atomic with respect to failure. Hsu et al. \cite{ref18} propose a programming model and runtime that adds persistence to existing multi-threaded C/C++ programs. Liu et al. \cite{ref43} propose Janus, a hardware-software co-design that parallelizes and pre-executes backend memory operations in an NVM system. Mnemosyne \cite{ref6}, NV-Heaps \cite{ref16}, and NVRAM transactions \cite{ref26} explored techniques for transactional interactions with PM. Checkpointing is often used for fault tolerance. Previous works \cite{ref22,ref23} for checkpointing on NVM focused on minimizing the checkpointing latency and bandwidth. 

Lin et al. \cite{ref31} adapt existing CPU persistency models for GPUs. DRAGON \cite{ref45} exploits NVM's capability for GPU through unified memory. GPM \cite{ref21} exploits UVA to map PM to GPU's address space to support fine-grain persistence to GPU kernels without the CPU's help.

Nalli et al \cite{ref13} propose WHISPER, a PM benchmark suite, and HOPS, a hands-off persistence system. HOPS tracks updates to PM in hardware and provides high-level ISA primitives for applications to express durability and ordering constraints separately. Themis \cite{ref27} exploits separate data paths to provide implicit ordering guarantees without using intervening fence instructions. DPO \cite{ref44} and PMEM-spec \cite{ref28} exploit a dedicated persist-path for PM $store$s and a regular path for PM $load$s.

\section{Conclusion}
\label{sec:conclu}
In this paper, we propose an adaptive approach (AGPM) for PM log updates' data path selection in the GPU. We first describe the observations we found in the comparison of the experimental results of the GPU kernels using a static data path selection strategy. We observe the presumption that it is more beneficial to use a non-temporal path for undo logging does not fit the GPU workloads. Besides, we also observe that the log updates possessing strong locality with the subsequent data updates are not always appropriate to use the temporal data path. We analyze the data locality among the GPU kernels' log updates and the subsequent data updates, as well as the data locality among the exclusive log updates, in the GPU's L1D and L2 caches. We translate these analyses into 8 reasons and use these reasons to guide the data path selection in our proposed solution. We implement L1- and L2-level AGPM buffers to record the statistics in GPU kernels' runtime execution, and we feed the statistic to the AGPM path selector in each SM to match with the reasons we concluded. Upon the matching results, AGPM adaptatively changes the data path for the log updates to PM. According to the experimental results, AGPM outperforms the state-of-the-art multi-data-path architecture to PM for different types of GPU workloads.

\section*{Acknowledgment}

TBD

\end{document}